\newif\ifTwoC
\global\TwoCtrue

\ifTwoC
\documentclass[twocolumn]{aastex631}
\else
\documentclass[manuscript,12pt]{aastex631}
\fi

\shorttitle{Passing stars \& paleoclimate}
\shortauthors{Zeebe and Hernandez}

\graphicspath{{./}{fig/}}

\newcommand{\giturl}{\url{github.com/rezeebe/orbitN}}
\newcommand{\zenurl}{\url{zenodo.org/records/8021040}}

\newcommand{\orbN}{{\tt orbitN}}
\newcommand{\airb}{{\tt airball}}
\newcommand{\ZBETa}{\mbox{\tt ZB18a}}
\newcommand{\HD}{\mbox{HD~7977}}
\newcommand{\SHD}{\mbox{$\tt S_{HD7977}$}}
\newcommand{\SEZ}{\mbox{$\tt S_{80}$}}
\newcommand{\dmin}{\mbox{$d_{\rm min}$}}
\newcommand{\MS}{\mbox{$M_{\Sun}$}}
\newcommand{\Ms}{\mbox{$M_{\star}$}}
\newcommand{\JT}{\mbox{$J_2$}}

\renewcommand{\v}[1]{\mbox{\boldmath$#1$}}
\renewcommand{\vr}{\mbox{$\v{r}$}}
\newcommand{\vv}{\mbox{$\v{v}$}}
\newcommand{\vSi}{\mbox{$\v{v}_{\Sun}$}}
\newcommand{\vvSi}{\mbox{$v_{\Sun}$}}
\newcommand{\eE}{\mbox{$e_{\Earth}$}}
\newcommand{\DeE}{\mbox{$\Delta e_{\Earth}$}}
\newcommand{\D}{\mbox{$\Delta$}}

\newcommand{\e}[1]{\mbox{$\times 10^{#1}$}}
\newcommand{\x}{\mbox{$\times$}}
\newcommand{\q}{\frac}
\newcommand{\sm}{\mbox{$\sim$}}

\newcommand{\alp}{\mbox{$\alpha$}}
\newcommand{\sig}{\mbox{$\sigma$}}
\newcommand{\tht}{\mbox{$\theta$}}
\newcommand{\tao}{\mbox{$\tau$}}
\newcommand{\rmx}{\mbox{$r_{\rm max}$}}
\newcommand{\rmxs}{\mbox{$r^2_{\rm max}$}}
\newcommand{\erf}{\mbox{erf}}
\newcommand{\erfinv}{\mbox{erf$^{-1}$}}

\newcommand{\gftL}{\mbox{($g_4$$-$$g_3$)}}
\newcommand{\sftL}{\mbox{($s_4$$-$$s_3$)}}
\newcommand{\MZ}{\mbox{${\cal M}_0$}}

\newcommand{\kms}{\mbox{km~s$^{-1}$}}
\newcommand{\ms}{\mbox{m~s$^{-1}$}}
\newcommand{\mspc}{\mbox{m~s$^{-1}$~pc$^{-1}$}}
\newcommand{\aud}{\mbox{au~d$^{-1}$}}

\newcommand{\beqn}{\begin{eqnarray}}
\newcommand{\eeqn}{\end{eqnarray}}

\newcommand{\myurl}{\url{www2.hawaii.edu/~zeebe/Astro.html}}
\newcommand{\npurl}{\url{www.ncdc.noaa.gov/paleo/study/26970}}

\begin{document}

\title{No influence of passing stars on paleoclimate 
reconstructions over the past 56 million years}

\author[0000-0003-0806-8387]{Richard E. Zeebe}
\affiliation{
SOEST, University of Hawaii at Manoa, 
1000 Pope Road, MSB 629, Honolulu, HI 96822, USA.
\correspondingauthor{Richard E. Zeebe}
\email{zeebe@soest.hawaii.edu} 
}

\author[0000-0001-7648-0926]{David M. Hernandez}
\affiliation{
Department of Astronomy, Yale University, Kline Biology Tower, 
219 Prospect St, New Haven, CT 06511, USA.
\email{david.m.hernandez@yale.edu} \\ \\ 
{{\rm Final Accepted Version,}
The Astronomical Journal}
}

\begin{abstract}
Passing stars (also called stellar flybys) 
have notable effects on the solar 
system's long-term dynamical evolution, injection of Oort cloud 
comets into the solar system, properties of
trans-Neptunian objects, and more. Based on a simplified solar 
system model, omitting the Moon and the Sun's quadrupole moment $J_2$,
it has recently been suggested that passing stars are also an important 
driver of paleoclimate before \sm{50}~Myr~ago, including a climate event 
called the Paleocene-Eocene Thermal Maximum (\sm{56}~Myr~ago). In contrast, 
using a state-of-the-art solar system model, including a lunar contribution 
and $J_2$, and random stellar parameters ($>$400 simulations), we 
find no influence of passing stars on paleoclimate reconstructions over 
the past 56~Myr. Even in an extreme flyby scenario in which the Sun-like 
star HD~7977 
($m = 1.07~M_\Sun$) would have passed within \sm{3,900}~au about 
2.8~Myr~ago (with 5\% likelihood), we detect no discernible change in 
Earth's orbital evolution over the past 70~Myr, compared 
to our standard model.
Our results indicate that a complete physics model is essential 
to accurately study the effects of stellar flybys on Earth's 
orbital evolution.
\end{abstract}

\keywords{
Solar System (1528)
Orbits (1184)
Celestial mechanics (211)
Solar neighborhood (1509)
}

\section{Introduction \label{sec:intro}}

Gravitational interactions due to stellar flybys
have a variety of effects on the solar system, including 
its long-term dynamical evolution, injection of Oort cloud 
comets into the solar system, properties of 
trans-Neptunian objects,
etc.\ \citep[e.g.,][]{oort50,rickman76,laughlin00,
malmberg11,liadams15,batygin20,zink20,
brownrein22,pfalzner24}. However, the vast majority of
flybys are weak and the likelihood of a
strong flyby is small, even over the solar system's
lifetime. Based on Gaia Data Release~3 (Gaia DR3), 
\citet{bailer22} estimated that the K7 dwarf Gl~710
($m = 0.7~\MS$) remains the closest known (future) 
encounter, with a median encounter distance of 
\sm{13,100}~au~=~0.0636~pc, to occur in about 1.3~Myr.
The second closest encounter was estimated to have
occurred \sm{2.8}~Myr ago with the G3 dwarf \HD\
($m = 1.07~M_\Sun$) and a median encounter distance of 
\sm{13,200}~au~=~0.0641~pc 
\citep[for more information on \HD, see, e.g.,][]
{delafuente22,bobylev22,dybczynski24,potemine24}.
Because most flybys are weak, the immediate consequences
for, e.g, the planetary orbits are small or negligible.
However, because of the solar system's chaotic dynamics,
small perturbations may have large-scale consequences in
the long run \citep[e.g.,][]{zink20,brownrein22,kaib24}.

The key question is, what is the 
relationship between the magnitude of the flyby 
perturbation and the magnitude of change in 
planetary orbits over a specified timescale 
(here $\mathcal{O}(10^8$~y)).
For instance,
\citet{kaib24} suggested that a flyby scenario in which the 
star \HD\ would have passed 2.8~Myr~ago within only 
\sm{3,900}~au \citep[5\% likelihood, see][]{bailer22},
would represent an important driver of paleoclimate before 
\sm{50}~Myr ago (for a discussion of causality, see
Section~\ref{sec:causl}). In other words, the magnitude of 
change in planetary orbits (in this case Earth's orbital 
eccentricity) in response to such a perturbation would
become significant over \sm{50}~Myr. It should be pointed
out that in addition to primary gravitational interactions,
the most accurate state-of-the-art solar system models already
include important secondary effects due to general relativity, 
the influence of the Moon (lunar contribution), the Sun's 
quadrupole moment \JT, and asteroids \citep[e.g.,][]
{laskar11,zeebe17aj,zeebe23aj}. 
Reliable predictions of orbital parameters based on 
these state-of-the-art models is limited to \sm{50}~Myr 
(without geological constraints) due to the solar system's 
chaotic behavior. So should stellar flybys be added
to accurate solar system models as an important secondary 
effect in multi-million year integrations? 

\def\tx{0ex}
\begin{table}[h]
\caption{Selected variables and notation. \label{tab:not}}
\vspace*{5mm}
\hspace*{-5mm}
\begin{tabular}{lllll}
\tableline\tableline
Variable & Unit  & Meaning   & Frame$^a$       & Note \\
\tableline   
\vv       & \kms & Stellar velocity vector     & LSR & [$v_1$ $v_2$ $v_3$] \\ [\tx]
$v$       & \kms & Stellar speed               &     & $v = |\vv|$         \\ [\tx]
\sig      & \kms & Stellar velocity dispersion &     & $^b$                \\ [\tx]
\vSi      & \kms & Sun's peculiar velocity     & LSR & $^c$                \\ [\tx]
$\v{X}_i$ & au   & Position solar system body $i$ &  ICS &                 \\ [\tx]
$\v{V}_i$ & \aud & Velocity solar system body $i$ &  ICS &                 \\ [\tx]
\tableline
\end{tabular} \\

\noindent {\small
$^a$Reference frame (coordinate system): LSR = Local Standard of Rest
\citep[see, e.g.,][]{mihalas81,binneytremaine08},
ICS = Integration Coordinate System for solar system bodies
(see Section~\ref{sec:num}). \\
$^b$For normally distributed velocities, $\sig =$ standard deviation;
$\sig^2 =$ variance. \\
$^c$Sun's ``apex velocity'' (depends on stellar type). \\[0ex]
}
\end{table}

The exact timescale here matters, i.e., it makes a difference
whether changes in orbital parameters due to flybys become 
significant over 50 or, say, 60~Myr. For example, a major 
climate event, called
the Paleocene-Eocene Thermal Maximum (PETM) occurred 56~Myr
ago and is widely considered the best analog for massive
carbon release to the Earth system \citep[e.g.,][]{zachos05natB}.
Earth's orbital configuration 56~Myr ago is relevant for
determining the exact age, chronology, and trigger of the PETM 
\citep{zeebelourens19}. Hence, passing stars could matter
for understanding
the PETM if the characteristic timescale of long-term 
flyby effects is 50~Myr, but not if the timscale is 60~Myr.
Importantly, note however that the finding of the PETM's 
onset coinciding
with a maximum in Earth's orbital eccentricity 
\citep[trigger mechanism, see] []{zeebelourens19} is 
primarily based on geologic data, and secondarily on 
astronomical calculations. Thus, including stellar encounters 
in astronomical models could potentially make a difference 
in the computations (although highly unlikely as we 
demonstrate here), but not in the data.
Below we show that including secondary 
effects such as a lunar contribution and \JT\ in 
solar system models \citep[not considered by][]{kaib24}
is critical to accurately determine 
the timescale over which flyby perturbations become 
significant in planetary orbital parameters. 

It should also be emphasized that the relationship 
between stellar flybys and orbital forcing on the one 
hand, and the relationship between orbital forcing and 
terrestrial climate change on the other, are two 
different and separate issues. The latter relationship
involves a complex, non-linear transfer function as the 
climate system response is strongly non-linear.
\ifTwoC\vspace*{30ex}\fi
The climate response is a separate problem
and does not enter the framework for the present
study \citep[for further information, see, e.g.,][]
{zeebekocken24esr}. The present study solely focuses  
on the relationship between stellar flybys and orbital 
forcing, which can be adequately described using the tools
presented in Sections~\ref{sec:meth} and~\ref{sec:num}.

\subsection{Causality \label{sec:causl}}

As mentioned above,
scenarios have been discussed in the literature in which 
the Sun-like star HD~7977 ($m = 1.07~\MS$)
passed relatively close to the solar system about 
2.8~Myr~ago \citep[e.g.,][]{bailer22}. 
\citet{kaib24} suggested that such a passage (if exceptionally
close) constitutes
``an important driver of paleoclimate'' before \sm{50}~Myr
ago, specifically focusing on the PETM \citep{zeebelourens19}. 
Considering causality, the
sequence of events begs the question, can a stellar close 
encounter at $t = -2.8$~Myr affect the PETM at $t = -56$~Myr? 
Of course not. However, it could potentially
affect our 
understanding of the PETM. Consider a large ensemble of
hypothetical histories (trajectories) of the solar system. 
When integrated starting from a point in the 
past toward the present, only those trajectories are
valid that satisfy the present state ($\v{X}_i$, $\v{V}_i$)
of solar system bodies at $t = 0$ compatible with present 
observations (for notation, see Table~\ref{tab:not}). 
Now select two of those trajectories.
One without (standard model), the other with a recent close 
encounter. Both trajectories are 
equally valid, only they have different histories, i.e., 
they diverge at some point in the past when looking at 
them from the present. For instance, as a result of solar 
system chaos, the trajectory including the close encounter 
would differ vastly from the standard model, say, beyond
80~Myr ago or so \citep[e.g.,][]{zeebe17aj}.
So the question is at what time in the past exactly do 
the trajectories diverge, and, would a different
trajectory, for example, change previous conclusions 
about orbital eccentricity forcing of Earth's climate 
at the PETM 56~Myr ago \citep{zeebelourens19}. 
In principle, it is straightforward to answer these 
questions because the physical laws involved are 
time-reversible, and, in practice, the numerical 
integrations are started in the present and the 
equations of motions are integrated
backward in time with a negative timestep. Thus, we
can simply explore the space of possible past trajectories
through ensemble integrations starting with valid
($\v{X}_i$, $\v{V}_i$) at $t = 0$ and include/exclude
recent close encounters, a series of distant stellar flybys,
etc. Importantly, it turns out that the decisive factor 
in properly answering the questions above is the complexity of
the solar system model employed in the integrations.

\section{Impulse \label{sec:imp}}

\begin{figure*}[t]
\begin{center}
\vspace*{-40ex}
\ifTwoC\vspace*{+4ex}\fi
\includegraphics[scale=0.7]{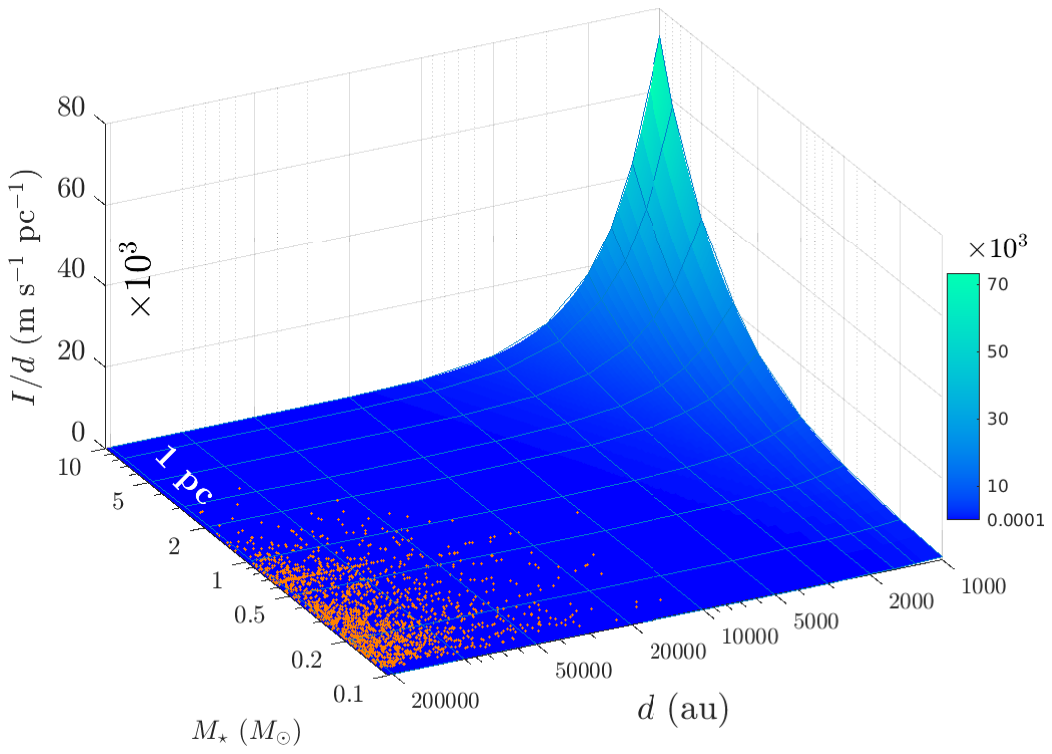}
\end{center}
\vspace*{-30ex}
\ifTwoC\vspace*{-6ex}\fi
\caption{
\footnotesize
Illustration of impulse gradient per unit mass 
(Eq.~(\ref{eqn:impgrad})) for stellar flybys
as a function of stellar mass ($\Ms$) and minimum passing 
distance ($d$). 
The upper $d$ limit shown is 1~pc$\ = 206,265$~au (see 
Section~\ref{sec:heisl}).
The colored surface uses a stellar speed $v = 50$~\kms.
Orange dots indicate example values for 1,800 flybys
based on randomly generated stellar parameters
(including $v$) after 
\citet{heisler87}, see Section~\ref{sec:meth};
1,800 flybys is an estimated number of stars passing 
the solar system within 100~Myr.
The figure's sole purpose is to illustrate 
the effect of different stellar parameters for very 
close, strong encounters (green-blue to green surface 
area) vs.\ average random flybys, which are weak (orange dots).
\label{fig:imp}
}
\end{figure*}

As an initial orientation for the effect of passing stars
on the solar system (or strength of the stellar perturbation)
one may consider the impulse ($I$). It appears 
natural to assume that the perturbation should somehow scale 
with the stellar mass ($\Ms$) of the passing star
and some inverse function of its 
speed ($v$) and minimum passing distance ($d$), aka impact
parameter. Indeed, to first order and for long stellar
passage times, the impulse per unit mass 
on the Sun may be approximated by
\citep[e.g.,][]{oort50,rickman76}:
\beqn
I = \q{2 G \Ms}{v d} \ ,
\label{eqn:imp}
\eeqn 
where $G$ is the gravitational constant; $I$ is in units
of \ms\ (SI).
To illustrate the effects of stellar mass and impact 
parameter on a solar system orbit, we use the parameter
$I/d$ (see Appendix~\ref{sec:impscale}):
\beqn
I/d = \q{2 G \Ms}{v d^2} \ .
\label{eqn:impgrad}
\eeqn
For example,
an exceptionally close encounter with a star of one solar
mass at $d = 1,000$~au and $v = 50$~\kms\ would generate
an impulse gradient per unit mass of $I/d \simeq 
7,300$~\mspc\ (see Fig.~\ref{fig:imp}).
Importantly, relative to 
an exceptionally close encounter, the average impulse 
gradient for 
the 1,800 randomly generated flybys shown in Fig.~\ref{fig:imp} 
is much smaller, i.e., $I/d = 0.42$~\mspc\
\citep[flybys after][see Section~\ref{sec:meth}]{heisler87}.
Briefly, the vast majority of flybys are weak.
Note that 1,800 flybys is an estimated 
number of stars passing the solar system within 100~Myr.

Note also that the equations given here to illustrate impulse 
and impulse gradient are for scaling purposes only.  
For our flybys, the inner planets complete multiple orbits 
on the stellar flyby's characteristic timescale. 
For this case, \citet{spurzem09} 
derive a scaling for the change in, e.g., eccentricity as 
$\propto d^{-2}$ for hyperbolic flybys, which is the same 
scaling as the impulse 
gradient (Eq.~(\ref{eqn:impgrad})); for parabolic flybys, 
\citet{spurzem09} derive a scaling $\propto d^{-3/2}$. Importantly, 
however, the results of the present paper are numerical (see 
below) and the impulse scaling does not affect our results or 
conclusions; they are merely provided to motivate our results.

\section{Methods: Flybys \label{sec:meth}}

To generate random stellar flyby parameters, we tested 
three options. (1) Following the method and parameters
from \citet{heisler87}, (2) taking parameters from 
\citet{garcia01}, and (3) using the \airb\ package 
\citep{brown24arb}.

\subsection{Stellar Flybys following Heisler et al. 
\label{sec:heisl}}

\subsubsection{Stellar type \label{sec:type}}

Define a quantity $S$:
\beqn
S = \sum_{i=1}^{20} n_i \sig_i \ ,
\eeqn
where $n_i$ is the number density per pc$^{-3}$ of stars of 
type $i$ ($N = 20$) and $\sig_i^2$ the variance of its velocity 
distribution ($\sig_i = $ dispersion in \kms, see Table~1 in 
\citet{heisler87}). Note that
$S$ is equivalent to a star flux (in SI units: 1/(m$^2$s)).
Cumulative and normalized, we can write:
\beqn
S_k = \sum_{i=1}^{k} n_i \sig_i / S \ ,
\label{eqn:sk}
\eeqn
where $0 < S_k < 1$.
Thus, for a uniformly distributed random variable $q$
($0 < q < 1$), Eq.~(\ref{eqn:sk}) can be used to assign a 
stellar type with $\sig_i$ and mass $m_i$.
The standard model
of \citet{heisler87} gives a mean encounter rate $f$
within $\rmx = 1$~pc of
\beqn
f = \sqrt{8\pi} \ S \ \rmxs \ = \rm 13.15~Myr^{-1} \ .
\eeqn
However, in our numerical simulations, stars were injected 
after certain time intervals with an average encounter rate
of \sm{18}~Myr$^{-1}$ (see Section~\ref{sec:numN}).
Note that $\rmx = 1$~pc was selected in the literature,
for example, to include aphelion distances of most
comets \citep{remymignard85} but has also been adopted
to study, for instance, impacts on planetary orbits 
\citep{kaib24}. We use the same value here for consistency. 
Furthermore, the impact of flybys on the solar system from 
encounters with $d \ \gtrsim \ 1$~pc is relatively small 
(see Fig.~\ref{fig:imp}).

\subsubsection{Stellar injection coordinates \label{sec:sic}}

Consider a sphere of radius $\rmx = 1$~pc around the 
Sun centered at the origin [0,0,0] (coordinates here
refer to the frame used in the numerical integrations,
see Section~\ref{sec:num}). Stars were injected
into the sphere at random positions
$\vr = [x \ y \ z]$, where
$x = \rmx \sin \tht \cos \phi$,
$y = \rmx \sin \tht \sin \phi$,
$z = \rmx \cos \tht$,
with polar angle $\tht$ ($-1 < \cos \tht < +1$)
and azimuth angle $\phi$ ($0 < \phi < 2\pi$), where
$\cos \tht$ and $\phi$ are each uniformly and 
independently distributed. Note that for a random,
isotropic distribution of \vr, the $z$~component 
is randomized (i.e, $\cos \tht$, not the angle \tht\ 
itself). 

\subsubsection{Stellar velocities}

At each position \vr\ on the sphere with radius \rmx,
stars were injected with velocity $\vv = [v_1 \ v_2 \ v_3]$,
where $v_3$ is anti-parallel to \vr\ and $v_1, v_2$
perpendicular to $v_3$. The general assumption for
the stellar velocity distribution is based on an
isotropic Maxwell-Boltzmann distribution
\citep[e.g.,][]{binneytremaine08,tremaine23}.
Importantly,
however, the velocity distribution for stars
originating from a given area $dA$ on the injection 
sphere is {\sl not} isotropic \citep{henon72}.

The distribution function (or normalized pdf)
for $v_3$ and a given stellar type may be written as
\citep{heisler87}:
\beqn
f_3(v_3) = \q{1}{\sig^2} \ 
                v_3 \exp(-v_3^2 / 2 \sig^2) \ ,
\eeqn
where $f_3(v_3) \cdot dv_3$ is proportional to the 
probability of finding
stars with velocities between $v_3$ and $v_3 + dv_3$.
To generate values for $v_3$, we sample the 
corresponding cdf:
\beqn
F_3(v_3) = 1 - \exp(-v_3^2 / 2 \sig^2) \ ,
\eeqn
using a uniformly distributed random variable 
$q$ ($0 < q < 1$), i.e.,
\beqn
v_3(q) = \left( -2 \sig^2 \ln (1 - q) \right)^\q{1}{2} \ .
\eeqn
The distribution function for $v_1$ and $v_2$ may be 
written as ($i = 1,2$):
\beqn
f(v_i) = \left( \q{1}{2\pi\sig^2} \right)^\q{1}{2}
                 \exp(-v_i^2 / 2 \sig^2) \ ,
\eeqn
with cdf:
\beqn
F_i(v_i) = \left[ 1 + \erf(v_i/\sqrt{2} \sig) \right]/2 \ ,
\eeqn
which we sample using uniformly distributed random variables
$q_i$'s:
\beqn
v_i(q_i) = \erfinv(2q_i-1) \cdot \sqrt{2} \sig \ .
\eeqn
Once the stellar velocities $\vv = [v_1 \ v_2 \ v_3]$
have been determined in coordinates relative to \vr\ 
(see above), they are transformed (rotated) into the 
solar system frame used in the numerical integrations 
(see Section~\ref{sec:num}).

\subsubsection{The Sun's peculiar velocity}

\citet{heisler87} did not take into account the Sun's 
peculiar velocity, \vSi, which we include here, loosely
following \citet{rickman08}.
Once the stellar velocity vector $\v{v}$ (randomized
directions), and its magnitude 
(speed) $v$ has been obtained as described above (relative 
to the Local Standard of Rest), we modify \vv\ to
obtain the final stellar velocity (encounter velocity) 
as follows. We assume a random relative 
3D orientation between the two vectors \vv\ and \vSi,
calculate the magnitude $|\vv - \vv_{\Sun}|$
and scale \vv\ by that factor \citep[for \vSi\ corresponding
to each stellar type, see][]{garcia01,rickman08}.
For example, for typical $v$ and \vvSi\ speeds of~40 and 
23~\kms, respectively, the procedure increases the 
final speed by about 4.4~\kms\ on average for random
$\vv$-$\vSi$ orientations (see Appendix~\ref{sec:vfav}).

\subsection{Stellar Flybys using parameters from Garcia et al.}

\begin{figure*}[t]
\begin{center}
\vspace*{-42ex}
\ifTwoC\vspace*{-10ex}\fi
\includegraphics[scale=0.7]{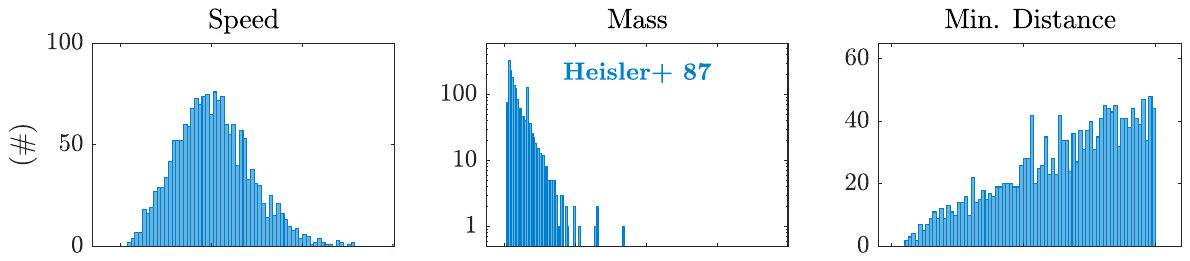}

\vspace*{-99ex}
\ifTwoC\vspace*{-22ex}\fi
\includegraphics[scale=0.7]{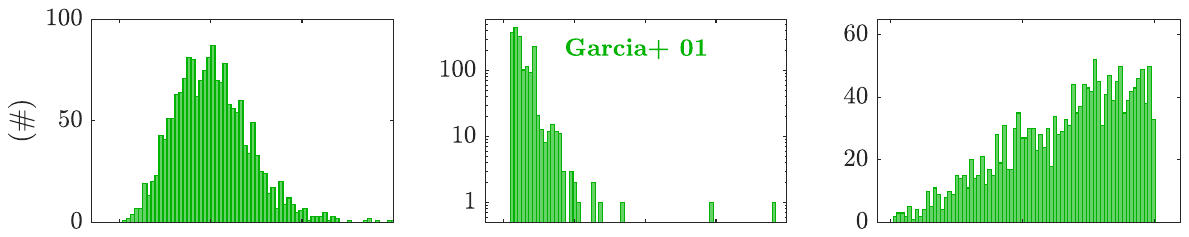}

\vspace*{-99ex}
\ifTwoC\vspace*{-22ex}\fi
\includegraphics[scale=0.7]{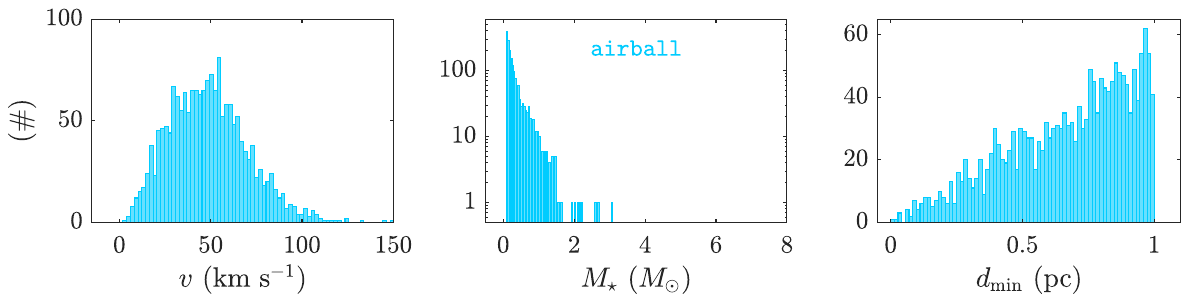}
\end{center}
\vspace*{-50ex}
\ifTwoC\vspace*{-10ex}\fi
\caption{
\footnotesize
Random stellar flyby parameters generated following 
\citet{heisler87}, \citet{garcia01}, and \airb\ (top, 
middle, bottom row). Shown are example histograms for 
1,800 flybys each. Note the logarithmic ordinate
for mass. 
\citet{garcia01} tends
to yield larger stellar masses at somewhat higher 
frequency, mostly due to a generally higher number 
density $n_i$ at a given mass (see text).
The largest \dmin\ considered here is 1~pc,
as stars are injected into a sphere around the Sun
with $\rmx = 1$~pc (see Section~\ref{sec:heisl}).
\label{fig:vmd}
}
\end{figure*}

For the second option of generating random stellar flyby 
parameters, we used \citet{heisler87}'s methods
but took the stellar parameters from \citet{garcia01},
see Table~8 and~1 in \citet{garcia01} and \citet{rickman08},
respectively. The velocity distribution is very similar 
to \citet{heisler87}, in terms of dispersion \sig,
most probable, and mean speed $\bar v$ (see 
Fig.~\ref{fig:vmd}). However, \citet{garcia01} tends
to yield larger stellar masses at somewhat higher 
frequency, mostly due to a generally higher number 
density $n_i$ at a given mass.

\subsection{Stellar Flybys using \airb}

For the third option, we used \airb\ \citep[a package 
for running and managing flybys,][]{brown24arb} to
generate parameters for our \orbN\ integrations.
For the stellar environment, we selected `Local Neighborhood',
a static environment representing the local neighborhood of 
the solar system. The velocity distribution is similar 
to \citet{heisler87} and \citet{garcia01} (see 
Fig.~\ref{fig:vmd}). However, the \airb\ environment
generates larger stellar masses at lower frequency
and appears not to include a significant population
of white dwarfs (cf.\ mass peaks at 0.65 and 0.9 \MS,
Fig.~\ref{fig:vmd} top and middle).

\section{Methods: Solar System Integrations \label{sec:num}}

\subsection{Basic integrator setup \label{sec:bset}}

Solar system integrations were performed following our 
earlier work \citep{zeebe15apjA,zeebe15apjB,zeebe17aj} with 
the integrator package \orbN\ \citep{zeebe23aj} using the 
symplectic integrator and Jacobi coordinates 
\citep{wisdom91,zeebe15apjA}. The simulations include 
contributions from general relativity, available as 
Post-Newtonian effects due to the dominant mass
\citep{saha94}. The 
Earth-Moon system was modeled as a gravitational quadrupole 
\citep{quinn91} ({\tt lunar} option), shown to be 
consistent with expensive Bulirsch-Stoer integrations 
(full dynamical separate Moon)
up to 63~Ma \citep{zeebe17aj}. Initial conditions for 
the positions and velocities of the planets and Pluto for 
\ZBETa\ (reference astronomical solution) were generated 
from the DE431 ephemeris \citep{folkner14} using the SPICE 
toolkit for Matlab. We had previously also tested the latest 
JPL ephemeris DE441 \citep{park21de}, which makes little 
difference for practical applications because the divergence 
time relative to \ZBETa\ (based on DE431) is \sm66~Ma and 
hence beyond \ZBETa's reliability limit of \sm58~Ma (based 
on geologic data, see below).
The integrations for \ZBETa\ 
\citep[timestep $\D t = -2$~days, avoiding numerical 
chaos, see][]{hernandez22}
include 10 asteroids 
\citep{zeebe17aj}, with initial conditions generated 
at \url{ssd.jpl.nasa.gov/x/spk.html}. The asteroids
were treated as heavyweight particles, subject to the same 
full interactions as the planets. Coordinates were obtained 
at JD2451545.0 in the ECLIPJ2000 reference frame and 
subsequently rotated to account for the solar quadrupole 
moment (\JT) alignment with the solar rotation axis 
\citep{zeebe17aj}. 
Earth's orbital eccentricity 
and inclination from \ZBETa\ is available at \myurl\ 
and \npurl. Importantly, we provide results from $-100$ 
to 0~Myr but caution that the interval from $-100$ to 
$-58$~Myr is unconstrained due to dynamical chaos
\citep{zeebelourens19}.

\subsection{Ensemble integrations \label{sec:numN}}

We generated 128 different sets of random stellar parameters for
each of the three options described in Section~\ref{sec:meth}.
Each set was integrated in parallel on the high performance 
computing cluster Derecho \citep{cisl23} for a total 
of $N = 384$ ensemble simulations. All other integration parameters 
were held constant (Section~\ref{sec:bset}). Stars were
injected after roughly equal time intervals, yielding
an average frequency of \sm{18} passing stars per million 
years, consistent with \citet{kaib24}.
In addition, we generated one hypothetical solution
\SHD, which includes randomly passing stars, as well as a 
(hypothetical) very close encounter with \HD\ at 
$t = -2.86$~Myr and $\dmin = 3,896$~au 
\citep[5\% likelihood, see][]{bailer22}.

\section{Results}

\subsection{Divergence time \tao\ and \HD}

\begin{figure*}[t]
\begin{center}
\vspace*{-35ex}
\ifTwoC\vspace*{-6ex}\fi
\includegraphics[scale=0.8]{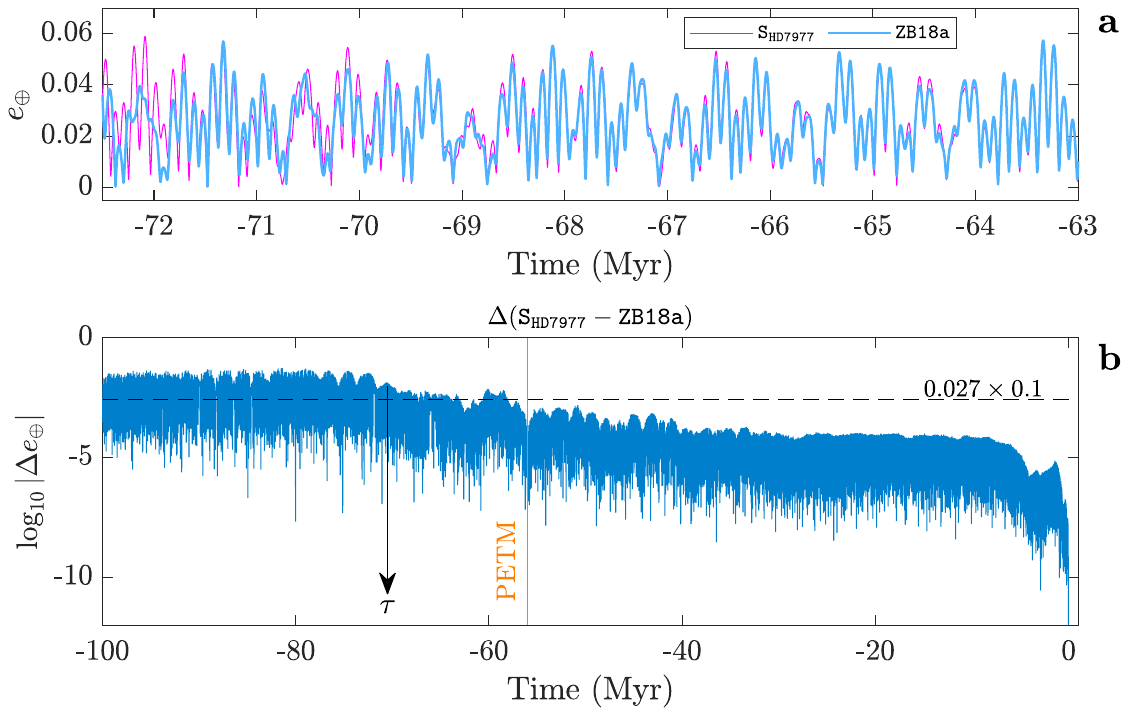}
\end{center}
\vspace*{-40ex}
\ifTwoC\vspace*{-6ex}\fi
\caption{
\footnotesize
Example of divergence time \tao. 
(a) Earth's orbital eccentricity, \eE, for solutions 
\SHD\ and our standard model \ZBETa. The setup for \SHD\ 
includes randomly passing stars, as well as a (hypothetical) 
very close encounter with \HD\ ($\dmin = 3,896$~au, 
$t = -2.86$~Myr). Notable differences between the two
solutions appear around $t = -70$~Myr.
(b) Difference in Earth's orbital eccentricity on a log-$y$ scale, 
$\log_{10}|\DeE|$, between solutions \SHD\ and \ZBETa\
over the past 100~Myr. 
The arrow indicates $\tao \simeq 70$~Myr (\tao\ is taken 
positive), when max$|\DeE|$ irreversibly crosses 
\sm{10}\% of mean \eE\ ($\sm0.027\x0.1$, dashed horizontal
line). Importantly, $\tau \simeq 70$~Myr is far beyond the PETM 
age ($\sm{56}$~Ma, vertical orange line).
\label{fig:tauHD}
}
\end{figure*}

The critical result of this study is the timescale
over which solutions that include passing stars differ 
from our standard model, or reference solution \ZBETa,
which does not include passing stars. We determine that 
timescale by using \tao, the time when max$|\DeE|$ 
irreversibly crosses \sm{10}\% of mean \eE; \tao\
is taken positive, as defined in \citet{zeebe17aj},
see Fig.~\ref{fig:tauHD}.
Note that defining \tao\ when max$|\DeE|$  crosses a given 
threshold once would frequently result in \tao's that 
are too short, because max$|\DeE|$ may drop again below the 
threshold beyond that time, where solutions show a good match
(see Fig.~\ref{fig:tauHD} and Section~\ref{sec:geo}). To avoid 
\tao's that are too short, our criterion includes the qualifier
``irreversibly''
\footnote{Strictly, crossing the threshold is not an irreversible 
operation, but appears irreversible on the $\sm\mathcal O(100)$ 
Myr time scale considered in this paper.}.
For $t > -\tao$, the solutions are practically
indistinguishable and hence flybys do not affect our 
understanding of any events within the time interval 
$[-\tao\ 0]$. For example, if $\tau > 56$~Myr, our
conclusions about orbital forcing of the PETM remain 
unchanged \citep{zeebelourens19}. For \SHD, $\tau 
\simeq 70$~Myr (Fig.~\ref{fig:tauHD}) and thus
even a hypothetical very close encounter at 
$t = -2.86$~Myr and $\dmin = 3,896$~au does not
alter our understanding of the PETM.

\subsubsection{Geologically discernible differences \label{sec:geo}}

The time at which the imprint of different \eE's
in a geologic archive would become discernible may slightly
differ from \tao. Different \eE's originate from 
different astronomical solutions and comparison with
geological data ultimately allows 
distinguishing/selecting between the solutions
\citep[see][]{zeebelourens19,zeebelourens22epsl}.
Earth's orbital eccentricity amplitude is 
modulated by the secular frequency term ($g_4 - g_3$), which 
leads to so-called very long eccentricity nodes (VLNs)
that can be detected in geologic archives
\citep[e.g.,][]{zeebelourens19,zeebekocken24esr}.
VLN intervals
show a weak amplitude of the short-eccentricity cycle  
(\sm{100}-kyr), with a recent period of \sm{2.4}~Myr.
VLNs occur in \ZBETa, for example, roughly around 
$-72, -70.5, -69$~Myr, and so on (see Fig.~\ref{fig:tauHD}a).
In \SHD, VLNs also occur roughly around $-$70.5 and $-$69~Myr,
but not around $-$72~Myr, where its 100-kyr amplitude is 
strong (for amplitudes and VLNs, see Fig.~\ref{fig:tauHD}a). 
In this case, the time
to distinguish between \ZBETa\ and \SHD\ based on geological
data using VLNs would be ca.\ $-$72~Myr, compared to ca.\ 
$-$70~Myr based on \tao. Note that the small \eE\ differences for
$t > -70$~Myr (Fig.~\ref{fig:tauHD}a) are unlikely 
to be detected in a geologic record.

\begin{figure*}[t]
\begin{center}
\vspace*{-35ex}
\ifTwoC\vspace*{-6ex}\fi
\includegraphics[scale=0.8]{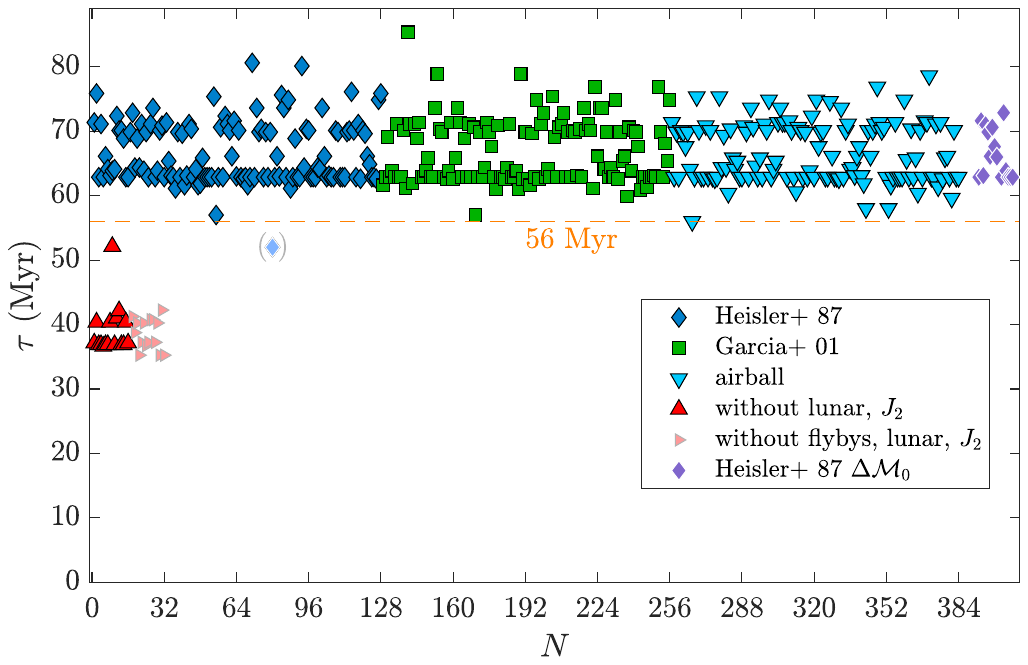}
\end{center}
\vspace*{-40ex}
\ifTwoC\vspace*{-6ex}\fi\caption{
\footnotesize
Divergence times, \tao's, for ensemble runs ($N = 3 \x 128$,
blue and green symbols)
relative to our reference astronomical solution (standard
model) \ZBETa\ (see Fig.~\ref{fig:tauHD}). 
Random stellar parameters for the three sets of
128 solutions are based on \citet{heisler87}, \citet{garcia01},
and \airb. All $\tau \geq 56$~Myr, except for the
symbol in parentheses ($\tau < 56$~Myr), indicating
one run (solution \SEZ) with a randomly generated 
(hypothetical), exceptional close encounter 
at $t \simeq -1.2$~Myr and $\dmin \simeq 699$~au,
for which there is no evidence in Gaia 
DR3 \citep[see text and][]{bailer22}.
Also shown are \tao's from runs including stellar flybys
but omitting the lunar contribution and \JT\ (red 
triangles, $\triangle$, $N = 16$), resembling the approach 
of \citet{kaib24}. Light red triangles ($\triangleright$, 
$N = 16$): no stellar flybys, no lunar contribution and 
\JT, but different initial positions for Earth 
\citep[see Fig.~4 in][]{zeebe23aj}.
Purple diamonds ($N = 16$): first 16~ensemble runs repeated
including random variations in the initial mean anomaly
\MZ\ of each body (see Section~\ref{dmz}).
For more information on \tao\ and solar system chaos, 
see \citet{zeebe23aj} and \citet{zeebekocken24esr}.
\label{fig:tauHGa}
}
\end{figure*}

\subsection{Ensemble integrations}

Of our 384 ensemble simulations, 383 show $\tau > 56$~Myr
(see Fig.~\ref{fig:tauHGa}), while a single solution
(\SEZ) shows $\tau < 56$~Myr. \SEZ\ represents
a simulation with a randomly generated (hypothetical), 
exceptional 
close encounter at $t \simeq -1.2$~Myr, $\dmin \simeq 
0.00339$~pc (699~au), and $\Ms = 0.23$~\MS. However, 
there is no evidence for such a close encounter in,
for instance, Gaia DR3, for which \citet{bailer22} 
found the G3 dwarf \HD\ to be the closest encounter 
in the past. Thus, our ensemble simulations 
(Fig.~\ref{fig:tauHGa}), as well as our solution 
\SHD\ (Fig.~\ref{fig:tauHD}), show that our understanding 
of the PETM, for instance, at $t = -56$~Myr is not 
affected by stellar flybys.

\begin{figure*}[t]
\begin{center}
\vspace*{-40ex}
\ifTwoC\vspace*{-6ex}\fi
\includegraphics[scale=0.8]{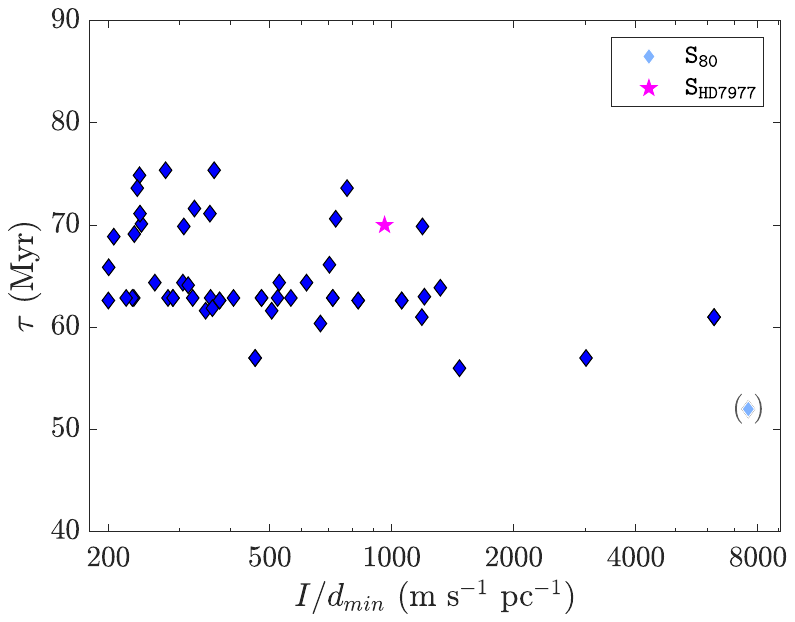}
\end{center}
\vspace*{-40ex}
\ifTwoC\vspace*{-6ex}\fi\caption{
\footnotesize
Divergence times, \tao's, vs.\ impulse gradient ($I/\dmin$)
shown for runs with strong close encounters $I/\dmin > 200$~\mspc\
that occur within the time interval $[-20\ 0]$~Myr. Note that 
flybys with $I/\dmin \ll 200$~\mspc\ are very frequent
(multiple per run) and exhibit no relationship with \tao\ 
at all (not shown, see text). Also note the logarithmic
abscissa.
\label{fig:tauI}
}
\end{figure*}

We had previously shown that a reduced physical model
which omits the lunar and \JT\ contributions
leads to substantially
shorter \tao's in solar system integrations
\citep[see Fig.~4 in][reproduced in Fig.~\ref{fig:tauHGa} here]
{zeebe23aj}. These simulations did not include stellar flybys 
but different initial positions for Earth. For the present 
study, we ran 16~additional simulations including stellar flybys 
and omitting the lunar and \JT\ contributions, which
resembles the approach of \citet{kaib24}. The results
show that regardless of whether initial conditions
are modified or flybys are included, a reduced physical 
model that ignores the lunar contribution and \JT\
enhances long-term instability and leads to substantially
shorter \tao's (see Fig.~\ref{fig:tauHGa}).
All of our simulations use an accurate and fast
implementation of general relativity (GR), i.e., 
Post-Newtonian effects due to the dominant mass,
following \citet{saha94}. In contrast, \citet{kaib24}
used a simplified central force modification to induce
GR precession (potential $\propto 1/(c^2r^2)$). 
However, given that our results for all \tao's (with
and without lunar and solar quadrupole) were obtained  
with the same GR implementation suggests that the GR 
model is of minor importance in this particular instance.

The small \tao\ for solution \SEZ\ with an 
exceptional close encounter may suggest a general
relationship between \tao\ and the strength of the 
flyby. However, dynamical chaos imposes at least two
restrictions on such a relationship. First, there is
a characteristic timescale over which flyby perturbations 
become significant in planetary orbital parameters,
i.e., around 50 to 60~Myr for the average flyby strength
considered here (see Section~\ref{sec:intro}). Thus, 
for a flyby solution to show significant differences
to the standard model, say at $t = -50$~Myr, the
encounter has to occur close to the present
(for a discussion of causality, see Section~\ref{sec:causl}).
Second, the exact long-term evolution of small 
perturbations is not predictable in detail due 
to chaos and often appears random \citep[see][]{zeebe23aj}. 
Thus, a correlation between \tao\ and the flyby strength
is only expected for strong encounters. 
Note that our \tao\ measures differences across a 
50-Myr timescale, which is different from a 1-Myr timescale 
\citep[see Fig.~3 in][]{kaib24}.
For strong encounters with impulse gradients 
$I/\dmin > 200$~\mspc\ that occur within the time interval 
$[-20\ 0]$~Myr, a weak relationship between \tao\ and 
$I/\dmin$ appears to be present (see Fig.~\ref{fig:tauI}).
For example, the smallest \tao\ is indeed associated with
the largest flyby impulse gradient (solution \SEZ).
However, for the vast majority of flybys (which are weak), 
the impulse gradient is in general a poor predictor 
for \tao, whose variability is dominated by dynamical 
chaos.

\subsection{Randomized initial conditions \label{dmz}}

In order to explore the possible parameter space of orbital solutions
resulting from dynamical chaos, randomized initial conditions (unique to each
solution) have been used in the past \citep[e.g.,][]{laskar11,
zeebe15apjB}. For otherwise identical integrations, unique sets of 
initial conditions ensure that no two ensemble members are alike.
Importantly, however, in the present case the integrations are {\sl not} 
identical because of different sets of random stellar parameters 
generated for each ensemble member, resulting indeed in 384 unique 
solutions (see Fig.~\ref{fig:tauHGa}). \cite{kaib24} varied
the initial mean anomaly (\MZ) of each body at $t_0$ by a random 
amount, generating an orbital shift between $\pm$2~cm (ca.\ 
$\pm1.3\e{-13}$~au). However, the magnitude of the
perturbation due to a typical stellar flyby (1,800 of which occur 
throughout the simulation, including close to $t_0$) is much larger.
Thus, randomizing the initial conditions for 
the current ensemble runs is unnecessary.

Nevertheless, to demonstrate the effect of random flybys vs.\ random 
initial conditions, we repeated the first 16~runs of our ensemble 
simulations \citep[following][]{heisler87}, as well as \SEZ, with 
small modifications. Note that \SEZ\ includes a randomly generated 
(hypothetical), exceptional close encounter at $t \simeq -1.2$~Myr. 
Relative to the reference initial conditions of \ZBETa, \MZ\ of each 
body (including asteroids) at $t_0$ was randomized,
resulting in small offsets within $\pm$2~cm.
The results show that $\Delta\MZ$ has no systematic effect on
\tao\ (all \tao's $>$60 Myr, Fig.~\ref{fig:tauHGa}, purple diamonds). 
The reduced \tao's 
(Fig.~\ref{fig:tauHGa}, red triangles) are due to reduced physics, 
not initial conditions. The modified and original \SEZ\ runs yielded 
identical \tao's (although the simulations themselves were not 
identical).

\section{Conclusions}

In contrast to \citet{kaib24}, we find no influence 
of passing stars on paleoclimate reconstructions over 
the past 56~Myr. The reason for the contrasting results
is the simplified solar system model employed by 
\citet{kaib24}, which omits the Moon (or any lunar
contribution) and \JT. Notably, it has been demonstrated 
previously that both the lunar contribution and \JT\ have a 
stabilizing effect, i.e., extend the divergence time \tao, on
a 100~Myr timescale (see Fig.~4 in \citet{zeebe23aj}),
which is confirmed here.
In order to draw accurate conclusions about the solar 
system's stability and chaotic behavior on a 100~Myr timescale, 
full state-of-the-art models need to be employed that consider 
all known secondary effects, including general relativity, 
a lunar contribution, \JT, and asteroids \citep{zeebe23aj,
zeebekocken25aj}.

Briefly, the reason for, e.g., the lunar contribution to have a 
stabilizing effect (extending \tao) is its long-term effect
on secular frequencies, specifically involving the secular 
resonance $m\gftL - \sftL$, where $m = 1,2$. Full state-of-the-art 
solar system models show a resonance transition from $m = 1$ 
to $m = 2$ around $t = -50$~Myr \citep[e.g.,][]{laskar11,
zeebelourens19,zeebelourens22epsl}. Omitting the lunar 
contribution tends to shift the timing of the resonance 
transition toward the present, thereby making the system 
less stable and shortening \tao. The details 
are complex, however 
\citep[see discussions in][]{zeebe23aj,zeebekocken25aj}
and deserve a proper, separate analysis, which is beyond the 
scope of this paper.

Running accurate state-of-the-art solar system models
that include all known secondary effects is computationally 
expensive. As a result, long-term studies on, e.g., 
Gyr-timescale tend to be based on simplified
solar system models, or the outer planets alone
\citep[e.g.,][]{zink20,brownrein22}.
One question to be addressed in future work is whether or 
not the results of such studies would be significantly
different using more complete solar system models.

\vspace*{5ex}

\begin{acknowledgments}
{\bf Acknowledgments.}
This research was supported by Heising-Simons Foundation Grants 
\#2021-2800 and \#2021-2803 (R.E.Z. and D.H.M) and U.S. NSF grant 
OCE20-34660 to R.E.Z. 
We thank the reviewers for suggestions, which improved the manuscript. \\
\end{acknowledgments}

\software{
          \orbN, \giturl; on Zenodo: \zenurl,
          \cite{https://doi.org/10.5281/zenodo.8021040}.
          }

\vspace*{5ex}

\appendix
\vspace*{-4ex}
\section{Impulse Scaling \label{sec:impscale}}
For the effect of a passing star on a comet, 
\citet{rickman76} derived an equation for the impulse 
per unit mass:
\beqn
I' = \q{2 G \Ms}{v d^2} \cdot r_c \ \sin \alp \ ,
\eeqn
where $r_c = |\v{r_c}|$ is the comet's typical distance 
from the sun ($\v{r_c}$ = heliocentric position vector)
and $\alp$ the angle between the vector
pointing from the sun to the passing star
and the plane perpendicular to $\v{r_c}$
\citep[see Fig.~3 in][]{rickman76}. For small
eccentricities, $r_c$ may be replaced by the 
semimajor axis $a$. Furthermore, for a given
$a$ and flyby geometry (which sets $\alp$), $I'$ 
scales with:
\beqn
\q{2 G \Ms}{v d^2} =  I / d \ ,
\eeqn
where $I$ is given by Eq.~(\ref{eqn:imp}).
We use $I/d$ for illustration in Fig.~\ref{fig:imp}
and for the relationship between strong encounters 
and \tao\ in Fig.~\ref{fig:tauI}. The parameter 
$2G\Ms/(v d^2)$ corresponds to \citet{kaib24}'s 
impulse gradient.

\section{Peculiar velocity contribution
\label{sec:vfav}}

In the following, we estimate the average peculiar velocity 
contribution for random orientations relative to the stellar 
velocity.
Let us denote the stellar velocity as $\vv = [v_1\ v_2\ v_3]$, 
and the Sun's peculiar velocity here as $\vv_{\Sun} = 
\v{u} = [u_1\ u_2\ u_3]$; $|\v{u}| = u$.
The final encounter speed is:
\beqn
v_f & = & \left[
        (v_1 - u_1)^2
     +  (v_2 - u_2)^2
     +  (v_3 - u_3)^2
      \right]^{\q{1}{2}} \ .
\eeqn
Without loss of generality, select a coordinate system
in which $\vv = [0\ 0\ v_3]$ and the transformed 
$v_3 > 0$; $|\v{v}| = v_3$. Also, use spherical coordinates
for \v{u}:
\beqn
v_f & = & \left[
        (u \sin \tht \cos \phi)^2
     +  (u \sin \tht \sin \phi)^2
     +  (v_3 - u \cos \tht)^2
      \right]^{\q{1}{2}} \ .
\eeqn
Furthermore,
$
u^2 \sin^2 \tht \cos^2 \phi + u^2 \sin^2 \tht \sin^2 \phi
  + u^2 \cos^2 \tht = u^2 \ ,
$
and hence:
\beqn
v_f = \left(
      v_3^2 + u^2 - 2 v_3 u \cos \tht
      \right)^{\q{1}{2}} \ ,
\eeqn 
with maximum and minimum values of $v_f = v_3 \pm u$.
To calculate an average $v_f$ for random relative
orientations, use $\zeta = \cos \tht$ as variable 
(see Section~\ref{sec:sic} for isotropy) and integrate:
\beqn
\overline{v_f} = 
      \q{1}{2} \int_{-1}^{1}
      \left(
      v_3^2 + u^2 - 2 v_3 u \ \zeta
      \right)^{\q{1}{2}} d \zeta \ .
\eeqn
Using $a = v_3^2 + u^2$ and $b = 2 v_3 u$,
integration yields:
\beqn
\overline{v_f} = -
      \q{1}{2} \cdot \q{2}{3b}
      \left[
      (a - b \ \zeta)^\q{3}{2}
      \right]_{-1}^{1} \ .
\eeqn
For example,
for typical $v$ and $u$ of~40 and 23~\kms, respectively, 
$v_f = 44.4$~\kms, i.e., the final speed increases by 
about 4.4~\kms\ on average.

\section{Justifying Wisdom--Holman \label{sec:jWH}}

\texttt{orbitN} utilizes the Wisdom--Holman method (WH), first introduced 
in \cite{wisdom91}. We demonstrate that the assumptions of this method 
hold even for extreme hyperbolic flybys considered in this paper. WH 
functions by splitting the orbital Hamiltonian ($H$) into a dominant piece 
described by Keplerian conic sections, including hyperbolic orbits, 
and a perturbation to this motion. Expressed in equations, $H = A + B$, where 
$|B|/|A| = \mathcal O(\epsilon)$, and $\epsilon \ll 1$ can be taken 
as the ratio of Jupiter's mass to the solar mass ($= 10^{-3}$) in the 
case of solar system studies. Compared to a usual solar system
integration, the flyby introduces new terms into $B$, say $B_F$, and we 
need to check that, 
\begin{equation}
|B_F|/|A|  \le \mathcal O (\epsilon) \ .  
\label{eq:assump}
\end{equation}
The flyby also introduces new terms into $A$, say $A_F$. A reasonable 
assumption is that $|A_F| \ge |A|$, and so we can ignore $A_F$ in what 
follows. Using notation from \cite{hernandez17}, we have,
\begin{equation}
B_F = \frac{G m_0 M_{\star}}{u_f} 
      - \frac{G m_0 M_{\star}}{r_{0f}} 
      - \sum_{0 < i < f} \frac{G m_i M_{\star}}{r_{if}} \ .
\end{equation}
Here, $\v{u}$ refers to Jacobi coordinates and $f$ is the flyby 
index, $m_0$ is the solar mass, and $r_{if}$ is the separation of 
star and body $i$.  Indices $0 < i < f$ refer to the planetary and 
asteroidal masses.  Eq. (\ref{eq:assump}) holds if 
$\mathcal O(M_{\star}/r_{0 f}) \le  \mathcal O(m_J/r_{0 J})$, with 
$J$ the index of Jupiter
and $m_J/r_{0 J} \simeq 2\e{-4}$~au$^{-1}$.  
For example, for the strongest non-hypothetical flyby (\HD, 5\% 
likelihood) we have $M_{\star} = 1.07 \ m_0$ and $r_{0 f} = 
3,900$~au, and hence $M_{\star}/r_{0 f} = 2.7\e{-4}$~au$^{-1}$.
For these values, we see, $\mathcal O(M_{\star}/r_{0 f}) = 
\mathcal O(m_J/r_{0 J})$, justifying our use of WH.

\bibliography{aas63593.bbl}{} 
\bibliographystyle{aasjournal}

\end{document}